\documentclass[%
 reprint,
 amsmath,amssymb,
 aps,
]{revtex4-1}

\usepackage{graphicx}
\usepackage{dcolumn}
\usepackage{bm}

\usepackage{subfigure}
\usepackage{xcolor}

\begin{document}


\title{Thermomagnetic properties and Debye screening for magnetized quark-gluon Plasma using the extended self-consistent quasiparticle model}

\author{Sebastian Koothottil}
\email{sebastian$_$dop@uoc.ac.in}
\author{Vishnu M. Bannur}%
\affiliation{%
Department of Physics, University of Calicut, Kerala-673635, India 
}%





\begin{abstract}
The thermomagnetic behavior of quark-gluon plasma has recently received a lot of attention. In this work we make use of the extended self-consistent quasiparticle model to study the thermodynamic  properties of magnetized (2+1) flavor quark-gluon plasma.  The system is considered as a non-interacting system of quasiparticles with masses depending on both temperature and magnetic field.  This allows to obtain the equation of state of the system and the other thermodynamic properties such as the speed of sound. We use the extended self-consistent model to obtain the magnetization and show that QGP has a paramagnetic nature. In addition, we study the pressure anisotropy and calculate the transverse pressure. The obtained anisotropic pressure may be used in hydrodynamic studies of magnetized QGP produced in heavy-ion collisions.  Finally we examine the screening properties of magnetized QGP in the longitudinal direction by calculating the Debye screening mass. 
\end{abstract}

\pacs{Valid PACS appear here}
\maketitle


\section{Introduction}
Quark-gluon plasma (QGP), the state of matter believed to have existed shortly after the big bang, has been successfully created in high energy collisions \cite{back1,*adams1,*adcox1,*arsene1}.  The charged ions can produce large magnetic fields reaching up to $eB\approx (1-15) m_{\pi}^2$ \cite{skokov3,zhong4} during off-central collisions.  Magnetic fields created in this manner may exist only for a short while but can be stationary during this time  \cite{tuchinstatic1, marasinghetuchinstatic, tuchin, fukushimastatic}. The theoretical tools used to study QGP need modifications to incorporate effects of external magnetic fields and there has been flurry of research activity in this area \cite{Karmakar2019,Kurian2018,Kurian2019, Kurian2019a, Mukherjee2018, Kurian2019b, Bandyopadhyay2016, Endrodi2013, Jamal2017, Levkova2014, Ghosh2019, catalysis1, catalysis2, Gusynin1996, Fukushima2013,chiral1,chiral2,chiral3, Ghosh2019a,Bandyopadhyay2019, Das2017,Avancini2019,Farias2014,Rougemont2016,Fayazbakhsh2011,Fayazbakhsh2010,Singh2018}.    Measurements at the LHC \cite{Collaboration2019}, along with those at RHIC \cite{Adam2019}, are capable of providing new insights that can constrain the theoretical modeling.  The equation of state has a significant impact on the evolution of QGP \cite{Bhadury2019}.   The study of the equation of state of magnetized QGP is relevant in the contexts of cosmology\cite{Grasso2001} and strongly magnetised neutron stars too \cite{Duncan1992,Broderick2000,Cardall2001,Rabhi2008,Rabhi2011}. The investigation of the behavior of magnetized QGP is, therefore,  of importance \cite{Avancini2018}.

     There are different approaches to studying the effect of magnetic fields on QGP \cite{Kurian2017,  Koothottil2019, Karmakar2019, Levkova2014, Farias2017}. In \cite{Koothottil2019}, we formulated the extended self-consistent quasiparticle model and studied the thermodynamics of magnetized 2-flavor QGP. 
    
    In \cite{Koothottil2019}, we developed the extended self-consistent quasiparticle model and studied the behavior of energy density, longitudinal pressure, and entropy density of magnetized $2-$ flavor QGP. In this work, we apply the extended quasiparticle model to magnetized $(2+1)$ flavor QGP. We use it to study the behavior of longitudinal pressure, speed of sound, magnetization, pressure anisotropy, and Debye screening.  We start with the study of thermodynamics by obtaining the equation of state and calculating the speed of sound in the medium. We then make use of our model to study the magnetic response of QGP by finding the magnetization and show that the system exhibits paramagnetic behavior. Using the calculated magnetization, we go on to study the anisotropy between longitudinal and transverse pressures caused by the magnetization acquired by the system along the field direction.  We bring out the dependence of transverse pressure on temperature and magnetic fields. Finally, we study the screening properties of magnetized QGP in the longitudinal direction by calculating the Debye screening mass.

     The paper is organized as follows. In section II, we summarize the extended self-consistent quasiparticle model discussed in \cite{Koothottil2019}. In section III, we apply this formulation for the case of (2+1) flavor QGP and obtain its equation of state and speed of sound in the medium.  Section IV involves the calculation of the magnetization of QGP. In section V we study the pressure anisotropy. Section VI deals with the calculation of the Debye screening mass of magnetized QGP. In section VI, we discuss the results obtained and conclude in section VII.    
\section{The extended self-consistent quasiparticle model}
In the extended self-consistent quasiparticle model, the thermal properties of interacting real particles are modeled by noninteracting quasiparticles with thermomagnetic masses\cite{Koothottil2019}.

  In the self-consistent quasiparticle model\cite{Bannur2007,Bannur2007a,Bannur2007b,Bannur2007c,Bannur2008,Bannur2012}, the thermal mass is defined to be proportional to the plasma frequencies as, 
   \begin{equation}
m_g^2 = \frac{3}{2} \omega_p^2 \;\;\;\text{and}\;\;\; m_q^2 = (m_0 + m_f)^2 + m_f^2.\label{eq:massplasmafreq}
\end{equation}

The plasma frequencies are calculated from the density dependent expressions\cite{Bannur2008} 
\begin{equation}
 \omega_p^2 = a_g^2 g^2\frac{ n_g}{T} + d_q^2 g^2 \frac{n_q}{T},\label{eq:plasmafrequencygluons}
 \end{equation}
  for gluons and,
 \begin{equation}
 m_f^2 = c_q^2 g^2 \frac{n_q}{T} ,\label{eq:plasmafrequencyquarks}
 \end{equation}
 for quarks. Here $n_q$ and $n_g$ are the quark and gluon number density, respectively. $g^2=4 \pi \alpha_s$ is the QCD running coupling constant. The coefficients $a_g$, $a_q$, $bq$ are determined by demanding that as $T\rightarrow \infty$, $\omega_p$ and $m_f$ both go to the corresponding perturbative results.   The motivation for choosing such an expression for plasma frequency is that the plasma frequency for electron-positron plasma is known to be proportional to $n/T$ in the relativistic limit \cite{Medvedev1999, Bannur2006}. Since the thermal masses appear in the expression for the density, we need to solve the density equation self-consistently to obtain the thermal mass. The thermal mass, in turn, may be used to evaluate the thermodynamic quantities of interest.  The results obtained have shown a good fit with lattice data even at temperatures near $T_c$ \cite{Bannur2012}.
  
     In the presence of magnetic fields, the energy eigenvalue values are given as Landau Levels, 
      \begin{equation}
  E_j=\sqrt{m^2+k_z^2+2 j q_f |eB|} \label{eq:energyeigenvalueslandaulevel}
  \end{equation}
 and the momentum integral is modified as  \cite{Fraga2012,Mizher2010, Chakrabarty1996, Bruckmann2017, Tawfik2016}, 
 
  \begin{equation}
  \int \frac{d^3k}{(2 \pi)^3}\rightarrow \frac{ q_f |eB| }{2 \pi}\sum_{j=0}^{\infty} \int \frac{dk_z}{2 \pi}\left(2-\delta_{0j}
  \right),\label{eq:dimensionalreduction}
  \end{equation} 
  where $f$ is the flavor index, and $q_f $ is the absolute value of the electric charge. 
The above equations modify the expression for number density.  
 The thermomagnetic mass for quarks is obtained by using the modified equation for number density in equation\eqref{eq:plasmafrequencyquarks} and solving the resulting equation self-consistently\cite{Koothottil2019}.
 
The expression for the number density of gluons remains unchanged in the presence of magnetic fields as gluons are chargeless and the thermomagnetic mass for gluons is obtained by solving equation\eqref{eq:plasmafrequencygluons} in a self-consistent manner.  Note that, even though the expression for gluon density remains unchanged in the presence of magnetic fields, they acquire a thermomagnetic mass through the quark number density.  Using the thermomagnetic mass, we can obtain the thermodynamics and study the thermomagnetic properties of magnetized QGP. 
 \subsection{Thermomagnetic Coupling}
  The only ingredient we need in order to make calculations is a thermomagnetic coupling, a coupling that incorporates the effect of both temperature and magnetic fields. To this end, throughout this work, we make use of the one-loop running coupling constant that evolves with both the momentum transfer and the magnetic field \cite{Ayala2018} as,
  \begin{equation}
  \alpha_s(\Lambda^2, \mid eB \mid) = \frac{\alpha_s(\Lambda^2)}{1 + b_1 \alpha_s(\Lambda^2) \log\left(\frac{\Lambda^2}{\Lambda^2+ \mid eB\mid}\right)},
  \end{equation}
  The one-loop running coupling in the absence of a magnetic field at the renormalization scale is given by,
  \begin{equation}
  \alpha_s(\Lambda^2)= \frac{1}{b_1 \log (\Lambda^2/\Lambda_{\overline{MS}}^2)},
  \end{equation}
  where, $b_1= (11 N_c-2 N_f)/12 \pi$ and following\cite{Karmakar2019}, $\overline{MS}=.176 GeV $ at $\alpha_s(1.5 GeV)=0.326$ for $N_f =3$.
  
  It is to be noted that the above thermomagnetic coupling has been obtained using the Lowest Landau Level approximation suitable in a strong magnetic field limit ($eB \gg T^2$). As explained in \cite{Ferrer2015}, in this limit, the coupling is split into terms dependent on the momentum parallel and perpendicular to the magnetic field separately.  The coupling dependent on the transverse momentum does not depend on the magnetic field at all. We are interested in how the system responds to magnetic fields, and so we make use of the longitudinal part of the coupling constant.  The coupling is obtained in the one-loop order, and so this may be appropriate only at very high temperatures. We use this coupling as an approximation, and so our results are bound to be qualitative. A two-loop thermomagnetic coupling, which includes the contribution from higher Landau Levels, is expected to give quantitatively reliable results.

 \section{Thermodynamics of (2+1)  flavor QGP in the presence of magnetic fields}
In this section, we study the thermodynamics of $(2+1)$ flavor QGP. We are interested in the thermomagnetic correction, and hence we will drop the pure-field contributions \cite{Dexheimer2013, Ferrer2010, Blandford1982, Menezes2009} from our calculations.  

 We focus primarily on the qualitative thermomagnetic behavior of magnetized QGP. So, the inclusion of the effects of dynamically generated anomalous magnetic moments\cite{Ferrer2010a}, as done in \cite{Chaudhuri2020, Strickland2012} is out of the scope of our work.  
 \subsection{Thermodynamic pressure}
 For quarks, the grand canonical potential, within the self-consistent quasiparticle model is, 
 \begin{widetext}
 \begin{align}
\frac{\Phi_q}{V}  =- P_{q}  =&{}-T \frac{g_f q_f | e B|}{2 \pi^2}\sum_{l=1}^{\infty} (-1)^{l-1}   \sum_{j=0}^{\infty} (2-\delta_{0j})\left[\frac{T}{ l^2} \frac{ m_{q_j} l}{T} K_1\left(\frac{ m_{q_j} l}{T}\right) + \int_{{T_0}}^T \frac{ d\tau}{\tau}\;  m_{q_j} \frac{\partial m_{q_j}}{\partial \tau} K_0\left(\frac{ m_{q_j} l}{\tau}\right) \right],
 \label{eq:quarkpressuremagfield}
\end{align}
\end{widetext}
where, 
\begin{equation}
m_{q_j}  =\sqrt{ m_q^2 + 2j \mid q_f eB\mid}. \label{eq:mqj}
\end{equation}
Here we have taken $\mu=0$. Note that in the self-consistent quasiparticle model, the grand canonical potential is not equal to $-KT \log \mathcal{Z}$, where $\mathcal{Z}$ is the grand partition function, due to the temperature dependence of masses. This is the reason why the expressions for pressure in equations \eqref{eq:quarkpressuremagfield} and \eqref{eq:Pressuregluonsmagfield} do not match the corresponding expressions for an ideal gas even though in the quasiparticle model the system is considered as an ideal gas with temperature-dependent masses. There is an extra term that ensures thermodynamic consistency, as shown in reference \cite{Bannur2007a}. The temperature dependence of mass in quasiparticle models has led to a whole lot of discussion about thermodynamic inconsistency problem and introduction of additional terms like $B(T)$ whose physical meaning is not obvious\cite{Gorenstein1995, Gardim2007}. The self-consistent quasiparticle model avoids this problem by starting with the expressions energy density and number density and calculating everything else from them.

The contribution from gluons is, 
\begin{align}
\frac{\Phi_g}{V} = - P_{g} =&{} -T \frac{g_f}{2 \pi^2}\sum_{l=1}^\infty \frac{1}{l^4} \Big[  T^3 \left(\frac{ m_g l}{T}\right)^2 K_2\left(\frac{ m_g l}{T}\right)\nonumber\\ &{}+\int_{T_0}^T \frac{d\tau}{m_g} \; \tau^3 \frac{\partial m_g}{\partial \tau}  \left(\frac{ m_g l}{\tau}\right)^3 K_1\left(\frac{ m_g l}{\tau}\right) \Big].\label{eq:Pressuregluonsmagfield}
\end{align}

Here, $T_0$ is some reference temperature, suitably chosen.

  \subsection{Velocity of sound}
 
  The velocity of sound is a fundamental quantity that is used in the description of hot QCD medium. The velocity of sound square $c_s^2$ is given by, 
  \begin{equation}
  c_s^2 = \frac{\partial P}{\partial \epsilon} = \frac{dP/dT}{d\epsilon/dT},
  \end{equation}
 where, $\epsilon$ is the energy density which can be obtained from pressure using the thermodynamic relation, 
 \begin{equation}
 \epsilon = T \frac{\partial P}{\partial T} -P.
 \end{equation}

 \section{Magnetization} 
 
Magnetization can be obtained from the Grand canonical potential $\Phi$.  
   \begin{equation}
   \mathcal{M} = -\frac{1}{V} \frac{\partial {\Phi}}{\partial(eB)}.\label{eq:magnetization}
   \end{equation}
  We confine our calculation to the region where $eB$ is greater than zero. Note that the equation for magnetization in the self-consistent quasiparticle model is not related to the partition function as in \cite{Bali2014}. This is because of the additional terms in  \eqref{eq:quarkpressuremagfield} and  \eqref{eq:Pressuregluonsmagfield},which ensure thermodynamic consistency.  

  Using equation \eqref{eq:quarkpressuremagfield}, we get, for quarks, 
  \begin{widetext}
  \begin{align}
  \mathcal{M}_q =&{} \frac{T g_f q_f}{2 \pi^2} \sum_{l=1}^\infty (-1)^l \sum_j ^{\infty} (2-\delta_{0j}) \Bigg\{  \left[\frac{T}{l^2} \left(\frac{m_{qj} l}{T}\right) K_1\left(\frac{m_{qj} l}{T}\right) + \int_{{T_0}}^T \frac{ d\tau}{\tau}\;  m_{q_j} \frac{\partial m_{q_j}}{\partial \tau} K_0\left(\frac{ m_{q_j} l}{\tau}\right) \right] \nonumber\\ 
  &{} -eB \left[\frac{T}{l^2} \left(\frac{m_{qj} l}{T}\right) K_0\left(\frac{m_{qj} l}{T}\right) \frac{\partial}{\partial(eB)}\left(\frac{m_{qj} l}{T}\right) -\frac{\partial}{\partial(eB)} \left(\int_{{T_0}}^T \frac{ d\tau}{\tau}\;  m_{q_j} \frac{\partial m_{q_j}}{\partial \tau} K_0\left(\frac{ m_{q_j} l}{\tau}\right) \right) \right] \Bigg\}.\label{eq:quarkmagnetization}
  \end{align}
  \end{widetext}
 Similarly, we obtain the expression for magnetization of gluons from equation \eqref{eq:Pressuregluonsmagfield} as,  
 \begin{align}
 \mathcal{M}_g =&{} \frac{T g_f}{2 \pi^2} \sum_{l=1}^\infty \Bigg[\frac{\partial}{\partial (eB)} \int_{T_0}^T \frac{d\tau}{m_g} \; \tau^3 \frac{\partial m_g}{\partial \tau}  \left(\frac{ m_g l}{\tau}\right)^3 K_1\left(\frac{ m_g l}{\tau}\right)\nonumber\\ {}{}{}&\;\;\;-T^3 \left(\frac{ m_g l}{T}\right)^2  K_1\left(\frac{ m_g l}{T}\right) \frac{\partial}{\partial(eB)} \left(\frac{ m_g l}{T}\right)\Bigg]
 \end{align}

  \section{Pressure Anisotropy}

There has been some discussion in the literature regarding the existence of a pressure anisotropy, and it has been suggested that the anisotropy is scheme dependent\cite{Ferrer2010,  Kohri2004, Martinez2003, Chaichian2000, Potekhin2012,  Ferrer2012, Dexheimer2013, Bali2014, Bali2013}.  In the $\phi$ scheme, the presence of magnetic fields breaks rotational symmetry due to the magnetization of the system in the direction of the magnetic field, resulting in a pressure anisotropy. Thus, in this scheme, the pressure has a transverse component different from the longitudinal component.

  The transverse pressure is related to the longitudinal pressure as, 
  \begin{equation}
  P_{T} = P - eB \cdot  \mathcal{M}. 
  \end{equation}
  Using equations \eqref{eq:quarkpressuremagfield} and \eqref{eq:quarkmagnetization} the contribution from quarks to transverse pressure becomes,
   
 \begin{widetext}  
 \begin{align}
  \frac{(P_T)_q}{T} = \frac{g_f  q_f  (eB)^2}{2\pi^2} \sum_{l=1}^\infty (-1)^l \sum_j ^{\infty} (2-\delta_{0j})\Bigg[\frac{T}{l^2} {}&\left(\frac{m_{qj} l}{T}\right) K_0\left(\frac{m_{qj} l}{T}\right) \frac{\partial}{\partial(eB)}\left(\frac{m_{qj} l}{T}\right)\nonumber\\ 
  &-\frac{\partial}{\partial(eB)} \left(\int_{{T_0}}^T \frac{ d\tau}{\tau}\;  m_{q_j} \frac{\partial m_{q_j}}{\partial \tau} K_0\left(\frac{ m_{q_j} l}{\tau}\right) \right) \Bigg]. 
  \end{align}
  \end{widetext}
  
  Similarly, for gluons, 
 
  \begin{widetext}
  \begin{align}
 \frac{( P_T)_g}{T}=&{} \frac{g_f }{2 \pi^2}\sum_{l=1}^\infty \frac{1}{l^4} \Bigg\{  T^3 \left(\frac{ m_g l}{T}\right)^2 K_2\left(\frac{ m_g l}{T}\right)+\int_{T_0}^T \frac{d\tau}{m_g} \; \tau^3 \frac{\partial m_g}{\partial \tau}  \left(\frac{ m_g l}{\tau}\right)^3 K_1\left(\frac{ m_g l}{\tau}\right)\nonumber\\ &- eB\Bigg[\frac{\partial}{\partial (eB)} \int_{T_0}^T \frac{d\tau}{m_g} \; \tau^3 \frac{\partial m_g}{\partial \tau}  \left(\frac{ m_g l}{\tau}\right)^3 K_1\left(\frac{ m_g l}{\tau}\right)-T^3 \left(\frac{ m_g l}{T}\right)^2  K_1\left(\frac{ m_g l}{T}\right) \frac{\partial}{\partial(eB)} \left(\frac{ m_g l}{T}\right)\Bigg] \Bigg\}
  \end{align}
  \end{widetext}

 \section{Longitudinal Debye Screening Mass} 
  
At the leading order, Debye screening mass parameterizes the dynamically generated screening of chromo-electric fields, due to the strong interactions of hot quantum chromodynamics \cite{Ghisoiu2015}.  Calculations of the higher-order contributions called magnetic Debye screening\cite{Arnold1995, Bonati2017} are beyond the scope of this work. The presence of an external magnetic field causes an anisotropy, and we study the Debye mass in the longitudinal direction.  The ability of QGP to shield out the electric potential can be measured in terms of the Debye screening length, which is the inverse of the Debye Mass $(m_D)$.  The conventional definition  for Debye mass can be obtained either from the small momentum limit of the gluon self energy\cite{Rebhan1993},\cite{Shuryak1978},\cite{Ghosh2018},\cite{Silva2014},\cite{Schneider2002} or the semiclassical transport theory\cite{Mrowczynski1999},\cite{Kurian2017},\cite{Yagi2005}.  In the zero magnetic field case, the Debye mass can be defined as, 
\begin{align}
m_D^2 = \frac{4 N_c}{T}  g^2 \int \frac{d^3 k}{(2 \pi^3)} f(\omega_k) \left(1-f(\omega _k)\right),\label{eq:Debyemassdefn}
\end{align}
where $f(\omega_k)$ are the quasi-gluon, quasi-quark/antiquark distribution functions with 
\begin{equation}
\omega_k= \sqrt{m(T)^2 + k^2}\label{eq:omega}.
\end{equation}
 In the self-consistent quasiparticle model, all the medium effects are captured by the thermal masses of the quasiparticles $m(T)$. The distribution functions (in zero chemical potential) are, 
\begin{align}
f_g(\omega_k) = \frac{1}{e^{\beta\omega_k} - 1}, \text{ and},\;
f_q(\omega_k^f) = \frac{1}{e^{\beta \omega_k^f}+1}, \label{eq:distfncts}
\end{align}
for gluons and quark/antiquark flavor $f$ respectively.

   Expression for the contribution of gluons to the Debye mass in zero magnetic field, is obtained using equation \eqref{eq:Debyemassdefn} with \eqref{eq:omega} along with the first relation in \eqref{eq:distfncts} as,
   \begin{align}
   {m^2_D}_g =\frac{6 g^2 T^2}{\pi^2}&\Bigg[\sum_{l=1}^\infty  \frac{1}{l^2} \left(\frac{lm_g}{T}\right)^2 K_2\left(\frac{lm_g}{T}\right)\nonumber\\
   &-2\sum_{l=2}^\infty \frac{l-1}{l^3} \left(\frac{l m_g}{T}\right)^2 K_2\left(\frac{l m_g}{T}\right)\Bigg].
   \end{align}
The contribution to the Debye mass from a single quark flavor at zero magnetic field is obtained using the quark distribution function  \eqref{eq:distfncts} in equation \eqref{eq:Debyemassdefn} 
\begin{align}
{m^2_D}_q &=\frac{6 g^2 T^2}{\pi^2} \sum_{l=1}^\infty \frac{(-1)^{l-1}}{l^2} \left(\frac{l m_q}{T}\right)^2 K_2 \left(\frac{l m_q}{T}\right).
\end{align}

In the presence of magnetic fields, the expression of Debye mass for gluons remains the same, and the dependence on magnetic fields are incorporated simply by replacing the thermal masses by thermomagnetic masses obtained earlier.  

The expression of Debye mass for quarks in the presence of magnetic fields can be obtained by replacing the thermal masses by thermomagnetic masses, changing the dispersion relation in accordance with equation \eqref{eq:energyeigenvalueslandaulevel}
and modifying the momentum integration according to \eqref{eq:dimensionalreduction} in equation \eqref{eq:Debyemassdefn}. This gives,
\begin{align}
{m^2_D}_q(eB) = \frac{3 \mid q_f B\mid g^2}{\pi} \sum_{j=0}^\infty \int_0 ^\infty  \frac{dk_z}{T} (2- \delta_{0j}) \times \nonumber\\
 f_q(\omega_{k_z }^j) \left(1-f_q(\omega_{k_z}^j)\right) ,\label{eq:magneticdebyemass}
\end{align} 
where, 
\begin{equation}
f_q (\omega_{k_z}^j) = \frac{e^{-\beta \omega^j _{k_z}}}{1+ e^{-\beta \omega^j _{k_z}}},
\end{equation}
and,
\begin{equation}
\omega_{k_z}^j = \sqrt{k_z^2 + m_{q_j}^2},
\end{equation}
with $m_{q_j}$ given in equation \eqref{eq:mqj}. 
Along with these, equation \eqref{eq:magneticdebyemass} can be simplified to,
\begin{align}
{m^2_D}_q (eB) = \frac{3 g^2 \mid q_f B\mid}{\pi^2} \sum_{l=1}^\infty (-1)^{l-1}  \sum_{j=0}^\infty \left(2 - \delta_{0j}\right) \nonumber \\  \left(\frac{ l m_{q_j}}{T}\right) K_1 \left(\frac{l m_{q_j}}{T}\right)
\end{align} 

\section{Results and Discussions}

  \begin{figure}
\begin{center} 
\includegraphics[scale=0.55]{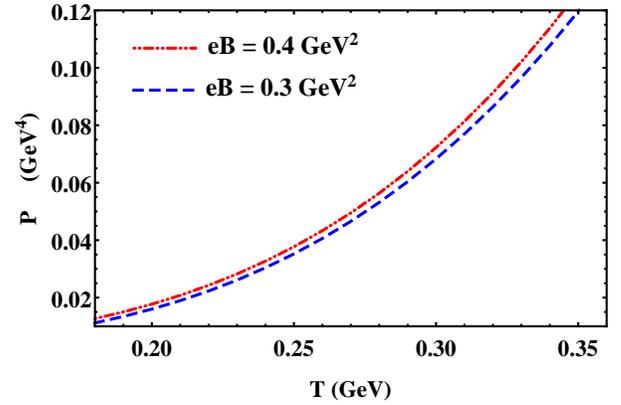}
\caption{Thermodynamic Pressure as a function of temperature different magnetic fields.}
\label{fig:pressure1} 
\end{center}
\end{figure}  
\begin{figure}
\begin{center}
\includegraphics[scale=0.55]{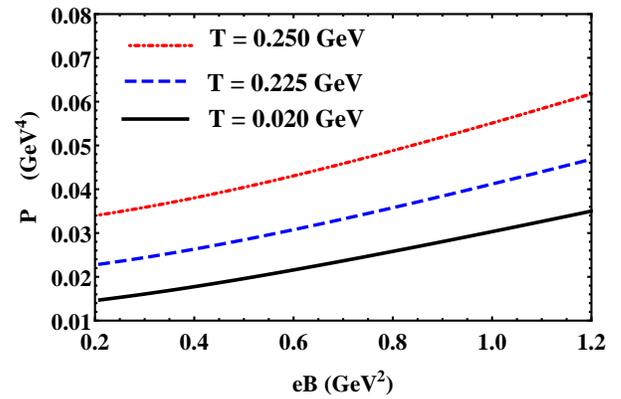}
\caption{Thermodynamic Pressure for different temperatures as a function of magnetic field.}
\label{fig:pressure2}
\end{center}
\end{figure} 
 \begin{figure}
\begin{center} 
\includegraphics[scale=0.55]{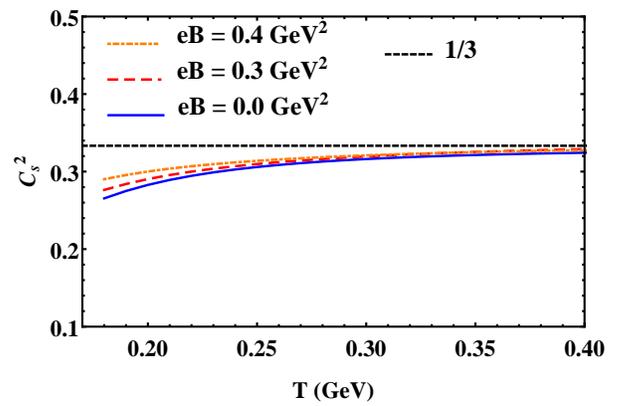}
\caption{Velocity of sound as a function of temperature for different values of magnetic fields}
\label{fig:CvsT} 
\end{center}
\end{figure}
  \begin{figure}
\begin{center} 
\includegraphics[scale=0.55]{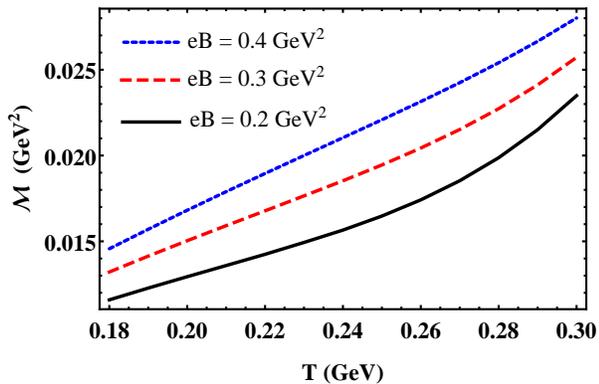}
\caption{Magnetization for different magnetic fields as a function of temperature.}
\label{fig:MvsT} 
\end{center}
\end{figure}
\begin{figure}
\begin{center}
\includegraphics[scale=0.55]{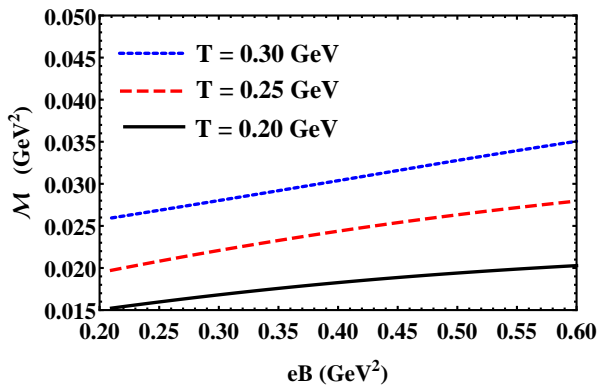}
\caption{Magnetization for different temperatures as a function of magnetic field.}
\label{fig:MvsB}
\end{center}
 \end{figure}
  \begin{figure}
\begin{center} 
\includegraphics[scale=0.55]{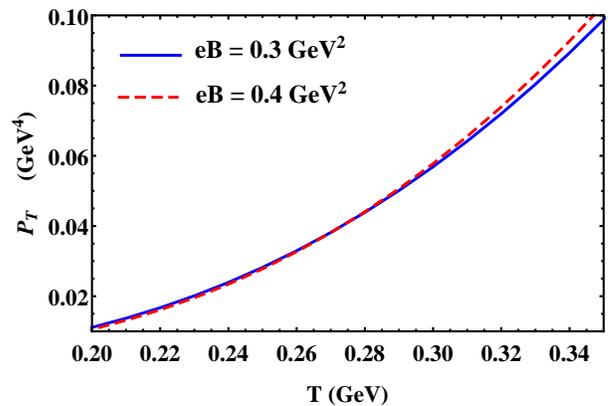}
\caption{Transverse pressure for different magnetic fields as a function of temperature.}
\label{fig:PtvsT} 
\end{center}
\end{figure}

   \begin{figure}
\begin{center}
\includegraphics[scale=0.55]{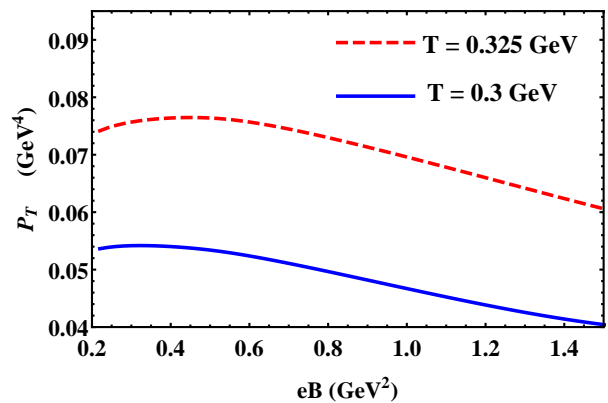}
\caption{Transverse pressure for different temperatures as a function of magnetic field.}
\label{fig:PtvsB}
\end{center}
\end{figure}
   \begin{figure}
\begin{center}
\includegraphics[scale=0.55]{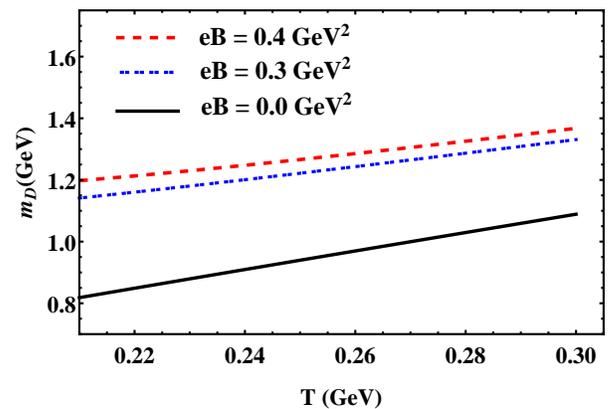}
\caption{Debye mass as a function of temperature for different magnetic fields.}
\label{fig:DebyemassvsT}
\end{center}
\end{figure}  
  \begin{figure}
\begin{center}
\includegraphics[scale=0.55]{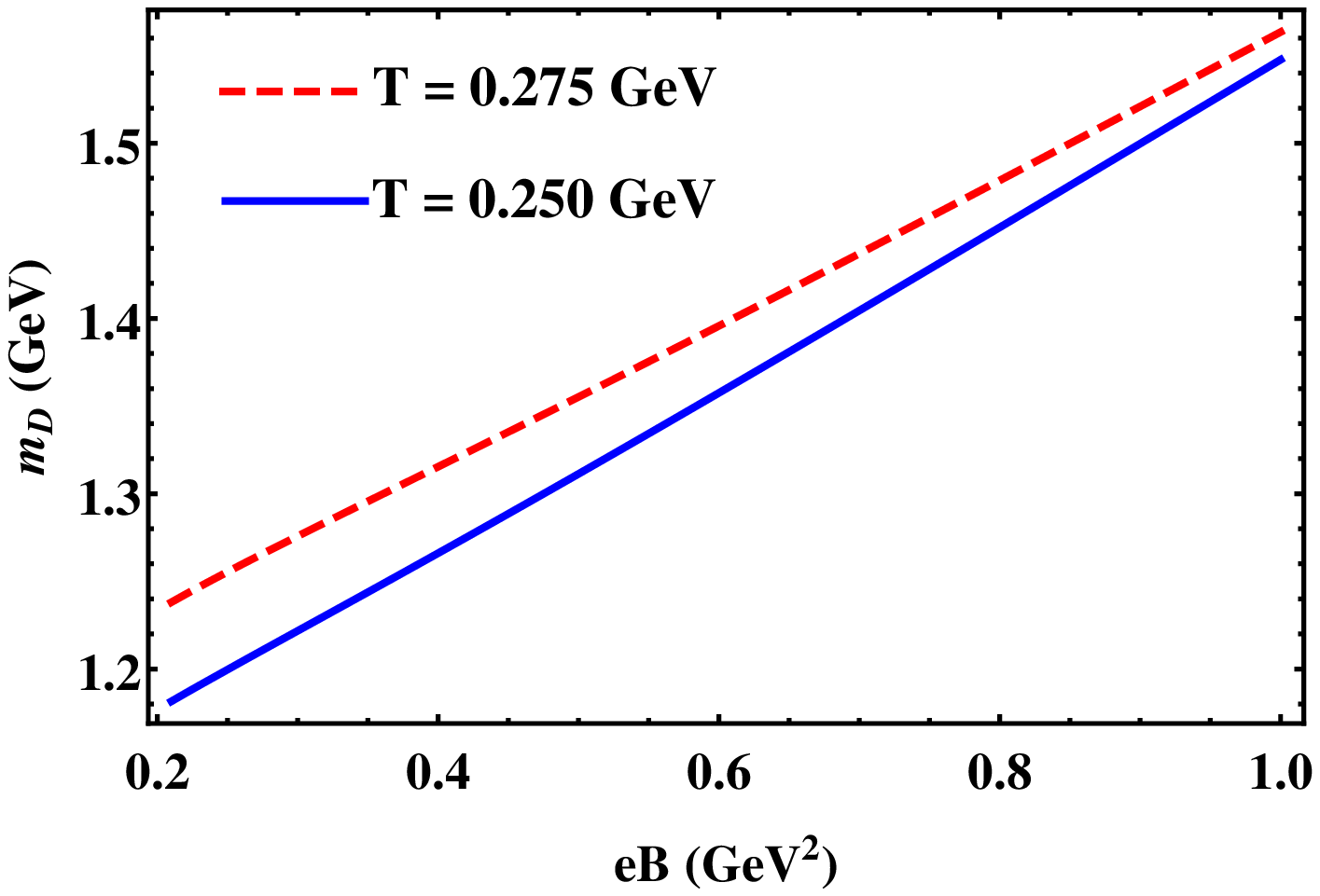}
\caption{Debye mass as a function of magnetic field  for different temperatures.}
\label{fig:DebyemassvsB}
\end{center}
\end{figure} 
   We have plotted the variation of longitudinal pressure with temperature, for different magnetic fields in Fig.\eqref{fig:pressure1}.  Fig. \eqref{fig:pressure2} shows the variation of pressure with the magnetic field for different temperatures. The increase in pressure with a magnetic field for at a given temperature, as seen in our equation of state, is consistent with lattice QCD results \cite{Bali2014}, perturbative QCD results\cite{Karmakar2019} and other works\cite{Kurian2017}. At this point, we do not make a quantitative comparison with the lattice data because the coupling constant may be reliable only at higher temperatures.  Besides, in our calculation, the effect of the anomalous magnetic moment is not included.

   In Fig.\eqref{fig:CvsT} we have plotted $C_s^2$ as a function of temperature, for different magnetic field values.  The speed of sound is seen to reach the Stefan-Boltzmann limit of $1/3$, asymptotically. This behavior is consistent with the behavior of $P/\epsilon$ in lattice QCD results \cite{Bali2014} and with the behavior of sound velocity in \cite{Kurian2017}.  
    
   The variation of magnetization with temperature for different magnetic fields is plotted in Fig.\eqref{fig:MvsT}. We see that the magnetization has a positive value for all values of temperature above $T_c$. This shows that QGP has a paramagnetic nature. The small deviation in the behavior of the graph for $eB = 0.2 GeV^2$ towards higher temperatures is because the contribution from even higher Landau Levels become relevant at this magnetic field.  In Fig.\eqref{fig:MvsB}, we have plotted the variation of magnetization with a magnetic field for different temperatures. It is seen that the magnetization increases with a magnetic field. The behavior of magnetization of QGP, as seen in our work, is qualitatively consistent with lattice QCD results \cite{Bali2014a}and with results from HTL perturbation theory  \cite{Karmakar2019}. We have included higher Landau Levels, whereas, in \cite{Karmakar2019} the Lowest Landau Level approximation has been used. We see that in our model, the contribution from higher Landau Levels cannot be neglected.
    
  In Fig.\eqref{fig:PtvsT}, we have plotted the variation of transverse pressure with temperature for different magnetic fields. In Fig.\eqref{fig:PtvsB}, we show the variation of transverse pressure with magnetic fields for different temperatures.  Since magnetization increases with temperature, the transverse pressure tends to decrease with an increase in the magnetic field. It may also go to negative values, indicating that the system may shrink in the transverse direction \cite{Bali2014a}. This behavior, too, is qualitatively consistent with the perturbative QCD results from \cite{Karmakar2019}, and lattice QCD results from \cite{Bali2014} and \cite{Bali2014a}.  
   
  Examination fo the screening effect in magnetized QGP using our model is studied by calculating the Debye mass. At $B = 0$, the Debye mass increases with temperature. With the increase in $B$, the Debye mass also increases. We have plotted the variation of Debye mass with temperature for different magnetic fields in \eqref{fig:DebyemassvsT}. The variation of the Debye screening mass with magnetic fields is plotted for different temperatures in \eqref{fig:DebyemassvsB}. The enhancement of Debye screening mass in presence of external magnetic field agrees with the finding in Ref.\cite{Bonati2017} using Lattice QCD simulations and in Refs.\cite{Bandyopadhyay2016, Alexandre2001, Bandyopadhyay2019}, and using perturbative calculations. Similar results are also obtained in \cite{Kurian2017} and \cite{Ghosh2018}.

\section{conclusion }
Summarizing, we made use of the extended self-consistent quasiparticle model to study the thermodynamics of ($2+1$)flavor Quark-Gluon Plasma. Importantly, we applied our model to investigate the magnetic response of the quark-gluon plasma. We then investigated pressure anisotropy in magnetized QGP. 
  
 We studied the thermodynamics of magnetized $(2+1)$ flavor QGP by plotting the pressure and sound velocity of magnetized QGP. The magnetic response of QGP was investigated by our model and the variation of magnetization with temperature and magnetic field proposed. We found that QGP has a paramagnetic nature. It has a small but positive magnetization at all temperatures above the transition temperature. We also noted that the presence of magnetization makes the system anisotropic, causing different pressures in directions parallel and perpendicular to the magnetic field. We evaluated the transverse pressure and plotted its variation with both magnetic fields and temperature. Finally, we studied the screening properties of magnetized QGP by examining the behavior of Debye screening mass in the longitudinal direction. We saw that the screening mass increases with magnetic fields.  Our results showed the same qualitative behavior as those obtained from Lattice QCD calculations and Hard Thermal Loop (HTL) perturbation theory approach and those obtained using other phenomenological models. The equation of state and anisotropic pressure calculated here can be used as an input for magnetohydrodynamic calculations and analysis of the elliptic flow of QGP formed in heavy-ion collisions.

   We see that the extended quasiparticle model is quite useful in studying various thermodynamic and thermomagnetic properties of the de-confined QCD matter in the presence of magnetic fields.  It is an advantage to this model that the higher Landau Level contributions can be incorporated without difficulty.  The present results could be improved with a two-loop order thermomagnetic coupling, which also incorporates the contributions from higher Landau Levels. Such a coupling and taking into account the anomalous magnetic moments would allow us to make quantitatively reliable predictions.

   It would be interesting to study the transport coefficients of magnetized QGP with the equation of state obtained using the extended self-consistent quasiparticle model. Another area where the model, with appropriate modifications, can be applied is QGP at finite temperature and density. Such a parametrization of the coupling strength would let us study the interior of a strongly magnetized neutron star using this model.

  \section{Acknowledgements} One of the authors S.K. thanks Manu Kurian for the helpful discussions. S.K. would like to thank University Grants Commission (UGC), New Delhi, for providing BSR-SAP fellowship
(F. No. 25-1/2014-15 (BSR)/7-180/2007 (BSR)) during the
period of research.

\bibliography{referenceaps2}

\end{document}